\documentclass[epj]{webofc}
\usepackage[varg]{txfonts}   
\usepackage{upgreek}
\woctitle{MESON2018 - the 15$^\textrm{th}$ International Workshop on Meson Physics}
\begin{document}
\selectlanguage{english}
\title{The onsets of deconfinement and fireball of NA61/SHINE}

\author{Dag Larsen\inst{1}\fnsep\thanks{\email{dag.larsen@cern.ch}} for the NA61/SHINE collaboration
}

\institute{Jagiellonian University, Krakow, Poland}

\abstract{
The NA61/SHINE experiment at the CERN SPS is pursuing a rich programme on strong interactions, covering the study of onset of deconfinement and aims to discover the critical point of strongly interacting matter by performing an energy and system-size scan at the full CERN SPS momentum range.
These scans of p+p, p+Pb, Be+Be, Ar+Sc and Pb+Pb have been mostly completed with Xe+La last year (more Pb+Pb to to taken this year).

Results from the different reactions are now emerging. As a surprise, some measurements did not scale smoothly. In particular, for the $\rm K^+/\piup^+$ ratio, Be+Be collisions behaved similarly to p+p (as superposition of nucleon collisions), while Ar+Sc was closer to Pb+Pb collisions.
This step can not be explained by onset of deconfinement, and may indicate that there is also a onset of fireball in relativistic heavy ion collisions.
A review of the results as well as possible interpretation will be presented.
The theoretical models (SMES, PHSD) describe onset of deconfinement at the heaviest system relatively well.
However, no model describes the behaviour of data at previously unmeasured collisions of light and intermediate size ions.
The onset of fireball is not described by any model.
}
\maketitle
\section{Introduction}
\label{intro}

The onset of deconfinement is a well established concept in strong interaction physics.
The search for it is also one of the goals of the current physics programme of the NA61/SHINE experiment by performing a energy and system-size scan at the CERN SPS.
However, the analysis of this scan has revealed phenomena that has properties of an onset, but can not be ascribed to the onset deconfinement.
This hits at an additional onset effect related to the size of the colliding systems.

\section{Onset of deconfinement}
\label{deconfinement}

\begin{figure}[h]
\centering
\begin{minipage}{0.47\linewidth}
\includegraphics[width=0.54\linewidth]{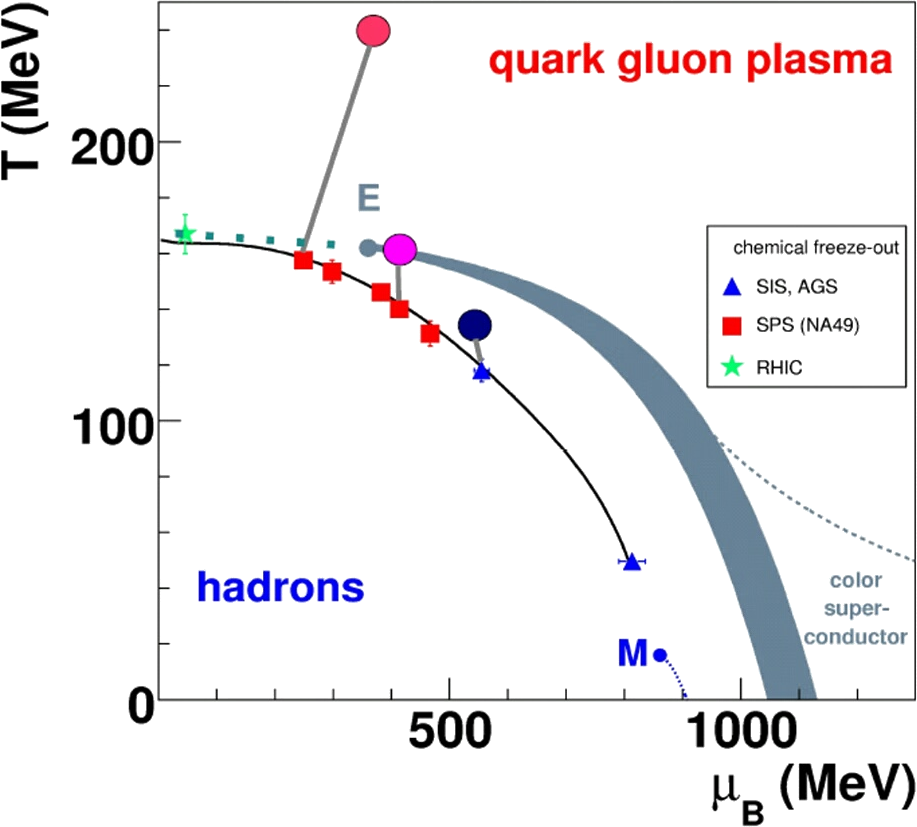}
\hspace{0.01\linewidth}
\includegraphics[width=0.42\linewidth]{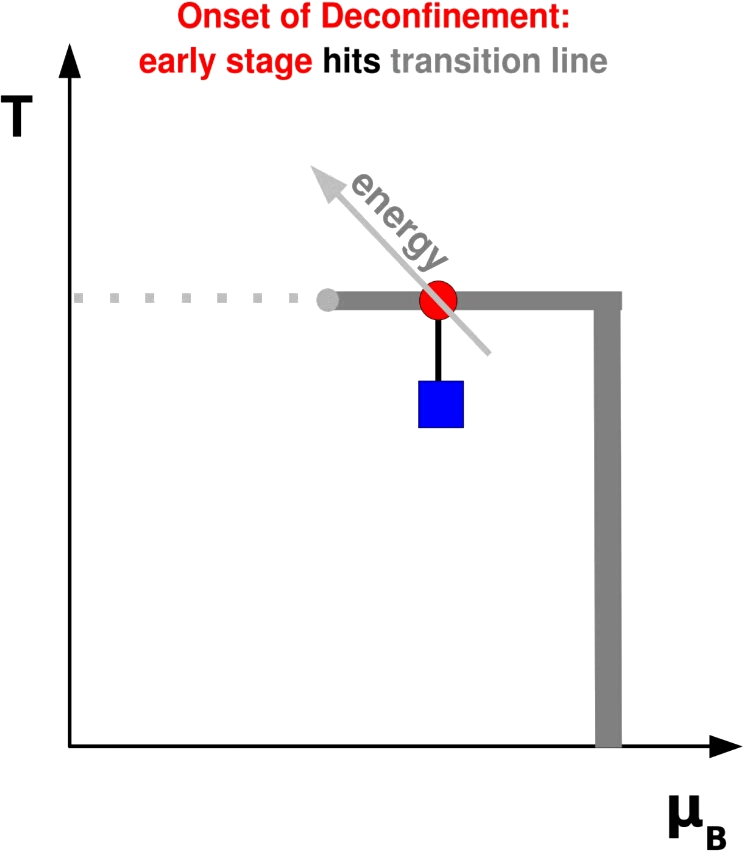}
\end{minipage}
\hspace{0.02\linewidth}
\begin{minipage}{0.47\linewidth}
\caption{The so-called phase diagram of strongly interaction matter {\emph(right)}. The onset of deconfinement takes place in the early stage when the transition line is crossed {\emph (left)}.}
\end{minipage}
\label{fig1}
\end{figure}

The creation of Quark--Gluon-Plasma (QGP) in A+A collisions begins with increasing collision energy $\sqrt{S_{NN}}$ as seen in Figure \ref{fig1}.
At lower temperatures and/or baryon chemical potential, strongly interacting matter is confined inside hadrons, while at higher temperatures and/or baryon chemical potential, strongly interacting matter is believed to be in a deconfined state in QGP.

\begin{figure}[h]
\centering
\includegraphics[width=0.28\linewidth]{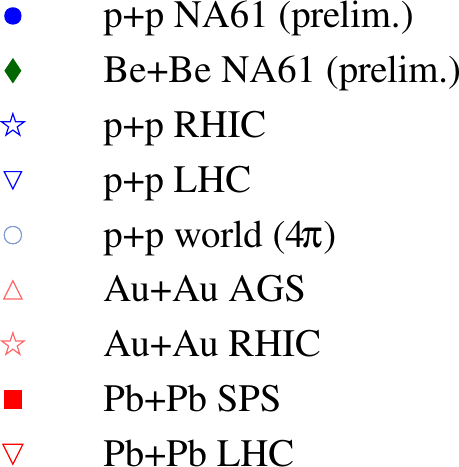}
\hspace{0.01\linewidth}
\includegraphics[width=0.3\linewidth]{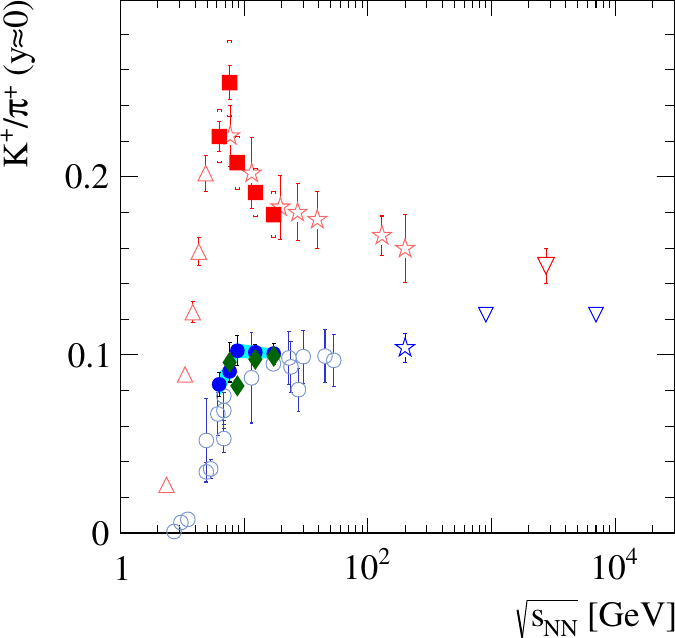}
\hspace{0.01\linewidth}
\includegraphics[width=0.3\linewidth]{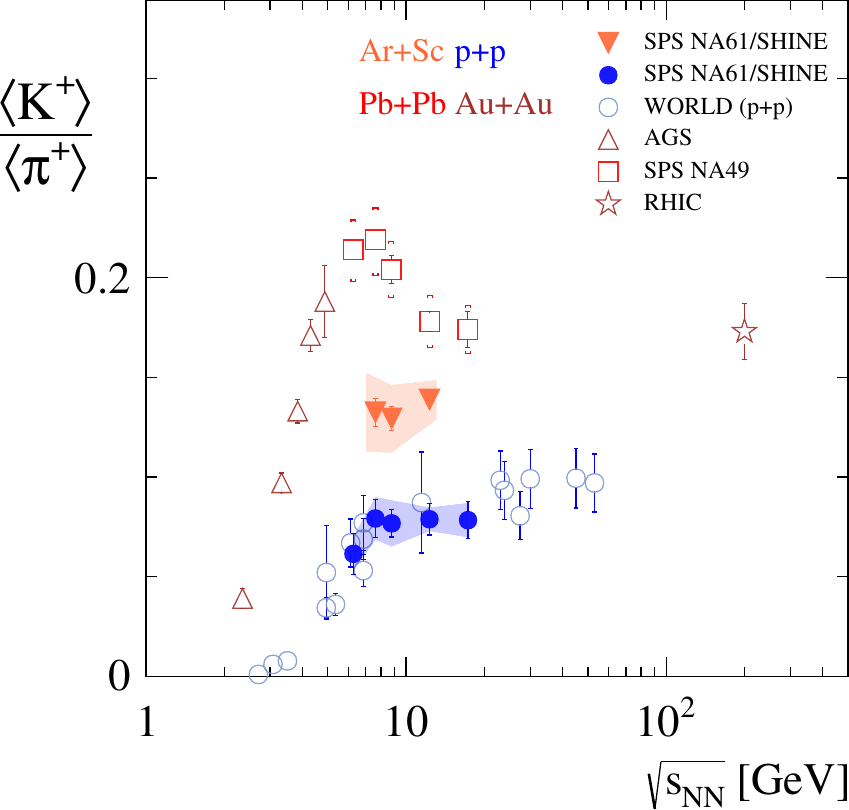}
\caption{Particle ratios for $\rm K^+/\piup^+$ {\emph(left)} and $\langle \rm K^+\rangle/\langle\piup^+\rangle$ {\emph(right)}. The horn structure is clearly visible for Pb+Pb collisions, while for p+p a step is observed. However, while Be+Be is close to p+p, Ar+Sc $\langle \rm K^+\rangle/\langle\piup^+\rangle$ is in between p+p/Be+Be and Pb+Pb.}
\label{fig2}
\end{figure}

There are several potential signatures of onset of deconfinement that can be searched for experimentally.
The particle ratio of $\rm K^+/\piup^+$, Figure~\ref{fig2}, is predicted by SMES~\cite{Gazdzicki:1998vd} to change rapidly around the transition to a deconfined state.
This effect is also known as the ``horn''.

\begin{figure}[h]
\centering
\includegraphics[width=0.28\linewidth]{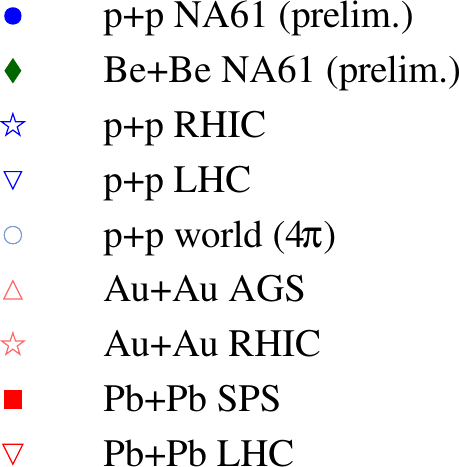}
\hspace{0.01\linewidth}
\includegraphics[width=0.3\linewidth]{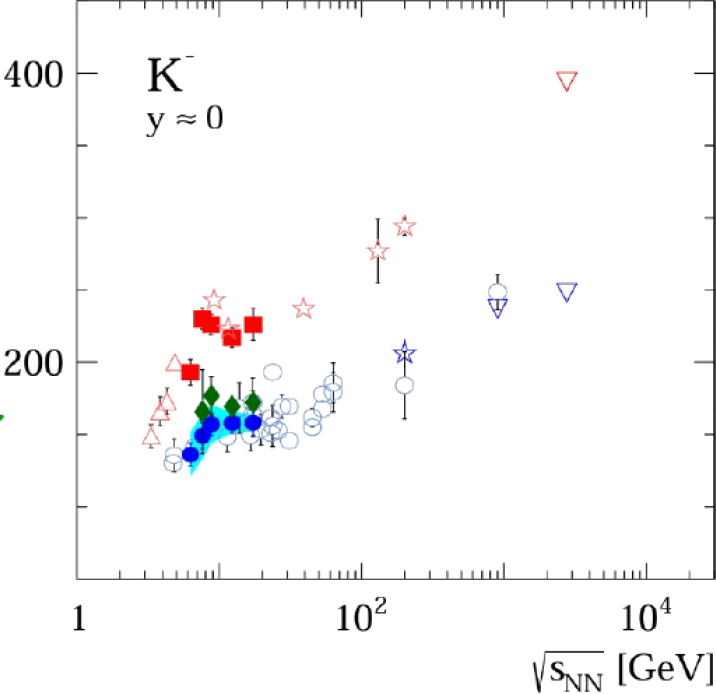}
\hspace{0.01\linewidth}
\includegraphics[width=0.3\linewidth]{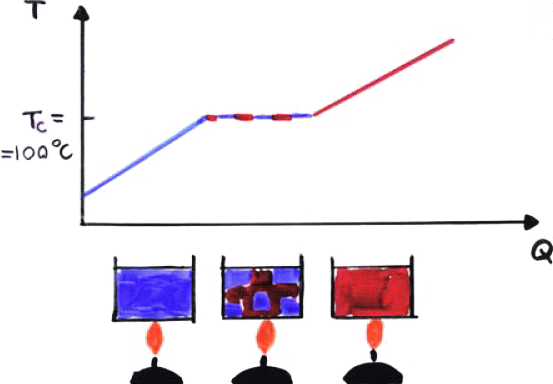}
\caption{The step is clearly visible for Pb+Pb collisions, but also seen in p+p collisions, as well as in Be+Be collisions which are close to p+p {\emph(left)}. The step-effect can be seen as an analogue to the temperature step for heating of water beyond the boiling point {\emph(right)}.}
\label{fig3}
\end{figure}

Another signature is the inverse-slope ``step'' in temperature versus collision energy.
This is also predicted by SMES as a signature of deconfinement.
Figure~\ref{fig3} shows current NA61/SHINE measurements of this structure.

\section{Onset of fireball}
\label{fireball}

\begin{figure}[h]
\centering
\begin{minipage}{0.48\linewidth}
\includegraphics[width=\linewidth]{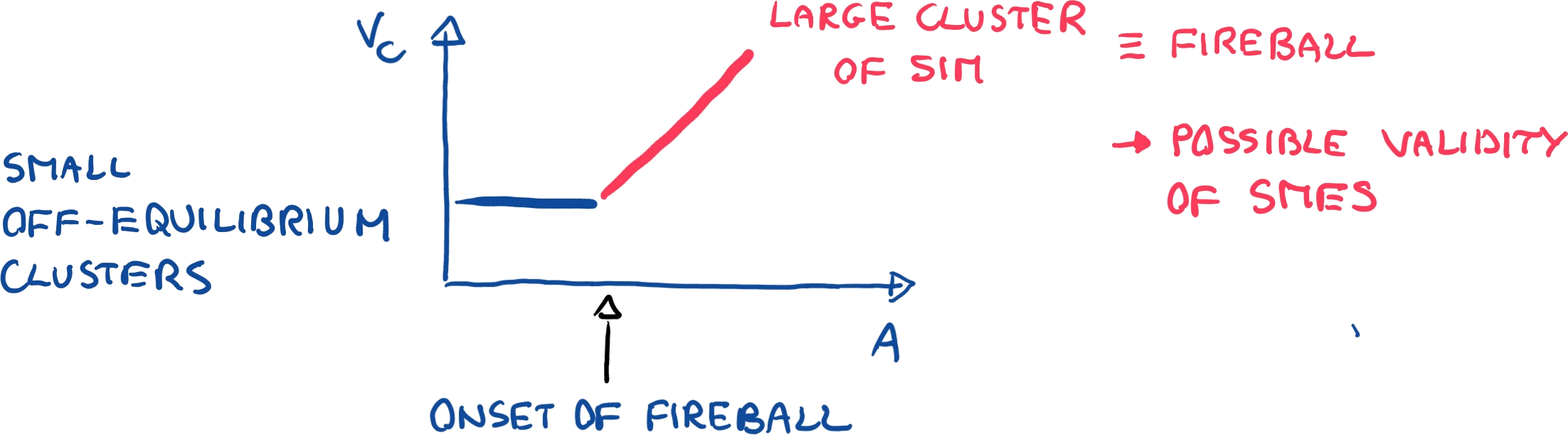}
\end{minipage}
\hspace{0.02\linewidth}
\begin{minipage}{0.48\linewidth}
\caption{Concept of onset of fireball. Until a certain mass number, only small, off-equilibrium clusters are created. Only when a mass number corresponding to onset of fireball will large clusters of strongly interacting matter be created.}
\label{fig4}
\end{minipage}
\end{figure}

Figure~\ref{fig2} showed that surprisingly the particle ratio for Be+Be was very close to p+p, while Ar+Sc was between p+p/Be+Be and Pb+Pb.
The naive expectation from the size of the colliding systems would be that Be+Be would be between p+p and Ar+Sc.
This may be a hint of a further onset effect, that may be referred to as ``onset of fireball''.
This is shown schematically in Figure~\ref{fig4}.
It is based on the assumption of beginning of creation of large clusters of strongly interacting matter (SIM) in nucleus-nucleus collisions with increasing mass number (A).

\begin{figure}[h]
\centering
\begin{minipage}{0.32\linewidth}
\caption{{\emph(Top.)} $\rm K^+/\piup^+$ ratio {\emph(left)} and multiplicity fluctuation {\emph(right)} versus number of wounded nucleons. A ``jump'' between light/intermediate and heavy systems is seen.
{\emph(Bottom.)} Possible impact of canonical suppression of $\langle N\rangle$ on $\rm K^+/\piup^+$ ratio {\emph(left)}, and canonical enhancement of $\omegaup$[N] on $\omegaup$[N] {\emph(right)}.}
\label{fig5}
\end{minipage}
\hspace{0.02\linewidth}
\begin{minipage}{0.64\linewidth}
\begin{minipage}{\linewidth}
\includegraphics[width=0.47\linewidth]{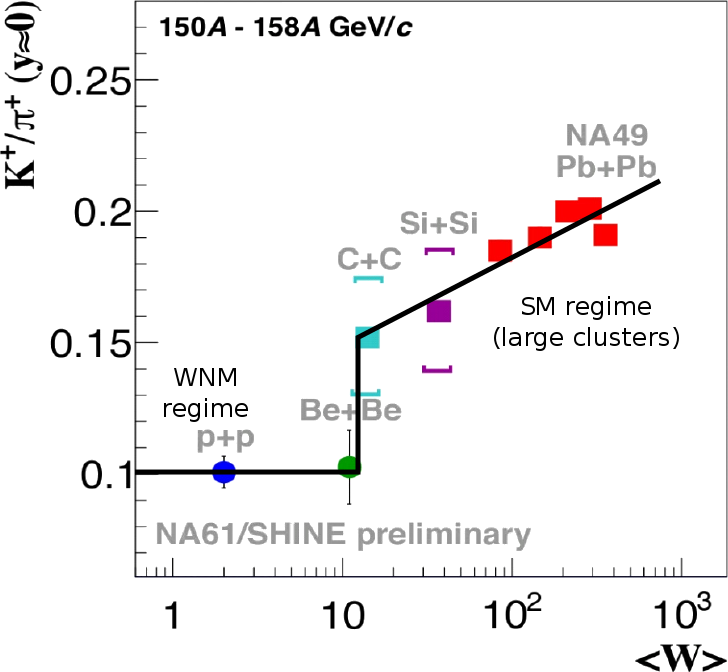}
\hspace{0.04\linewidth}
\includegraphics[width=0.47\linewidth]{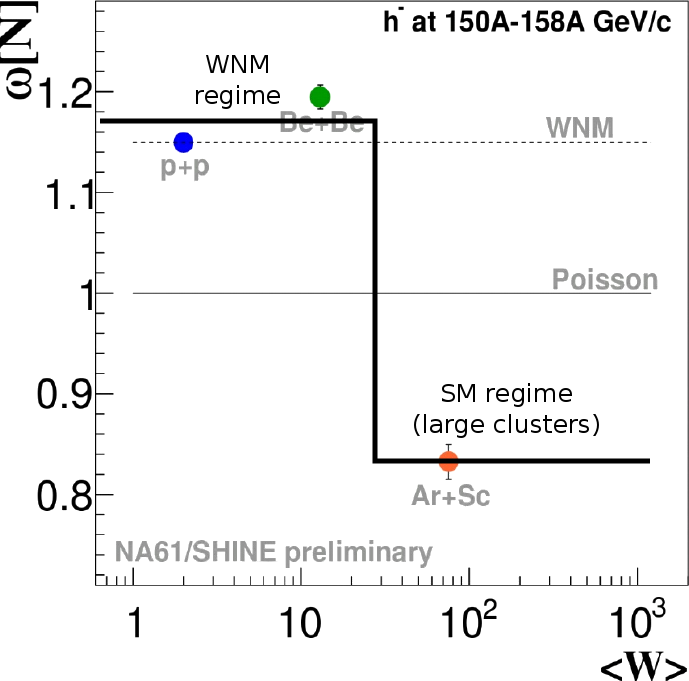}
\end{minipage}
\vskip 0.3cm
\begin{minipage}{\linewidth}
\includegraphics[width=0.47\linewidth]{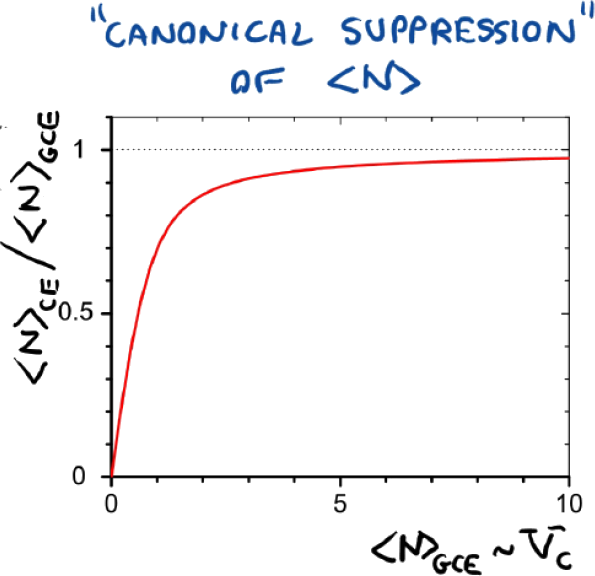}
\hspace{0.04\linewidth}
\includegraphics[width=0.47\linewidth]{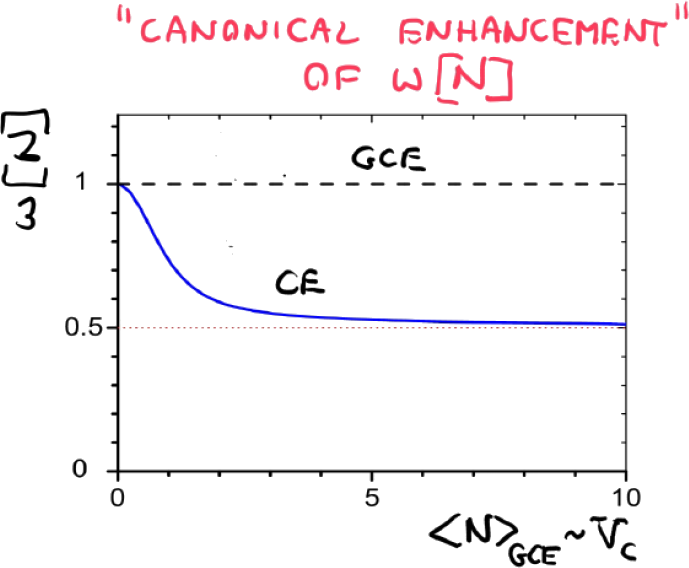}
\end{minipage}
\end{minipage}
\end{figure}

One may consider the $\rm K^+/\piup^+$ ratio and multiplicity fluctuation versus number of wounded nucleons.
These observables change rapidly when moving from light (p+p/Be+Be) to intermediate/heavy systems, as seen in Figure~\ref{fig5}.
The heavy systems are close to statistical model predictions for large volumes.
These observations may be consistent with a beginning of creation of large clusters of strongly interacting matter, {\emph i.e.} onset of fireball.


It may be possible to interpret these results in the context of the statistical model in terms of ideal Boltzmann gas and grand canonical ensemble versus canonical ensemble.
This would yield a canonical suppression of $\langle N\rangle$ that may explain the jump for the $\rm K^+/\piup^+$ ratio, and a canonical enhancement of w[N] that could explain the jump for $\omega$[N].
However, it is not clear if this can fully explain these effects.

\section{Possible interpretations of the onset of fireball}
\label{versus}

A possible interpretation is the percolation approach~\cite{Baym:1979etb}.
With increasing mass number A, the  density of clusters (strings, partons, {\emph etc.}) increases.
Thus the probability to overlap many elementary clusters may rapidly increase with A,  which may point to the validity of percolation models.
However, this approach does not explain equilibrium models of large clusters.

Another possible interpretation may be Ads/CFT correspondence~\cite{Maldacena:1997re}.
Ads in gravity is the formation of a black hole horizon (information trapping surface) that takes place when critical values of model parameters are reached.
CFT in QCD is that only starting from a sufficiently large nuclear mass number the formation of trapping surface in A+A collisions is possible.
This may be seen as an analogue to onset of fireball.

\section{D$^0$ as signal of deconfinement}
\label{d0}

\begin{figure}[h]
\centering
\begin{minipage}{0.32\linewidth}
\includegraphics[width=\linewidth]{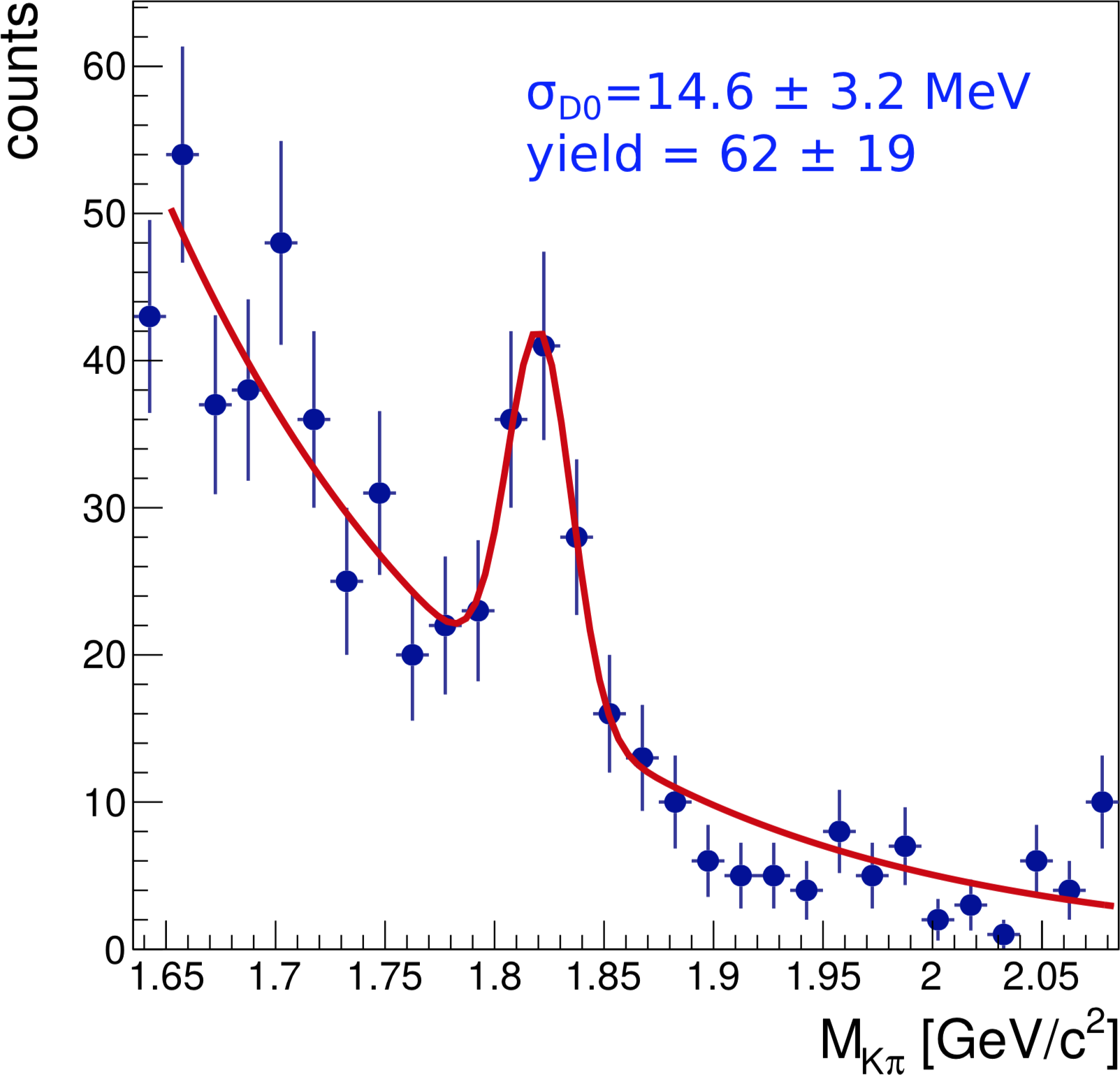}
\end{minipage}
\hspace{0.02\linewidth}
\begin{minipage}{0.64\linewidth}
\caption{First direct measurement of D$^0$ at SPS energies.}
\label{fig6}
\end{minipage}
\end{figure}

NA61/SHINE is undertaking open charm measurement programme with new Vertex Detector (VD)~\cite{d0}.
The charm yields expected to be different in a confined and a deconfined matter
So far pilot data has been taken, where a first direct observation, Figure~\ref{fig6}, of D$^0$ at SPS energies was seen.
Later this 2018 higher statistics will be taken to perform a more precise measurements.
An upgraded VD is expected to be introduced in 2021.

\section{Summary}
\label{summary}

\begin{figure}[h]
\centering
\begin{minipage}{0.32\linewidth}
\includegraphics[width=\linewidth]{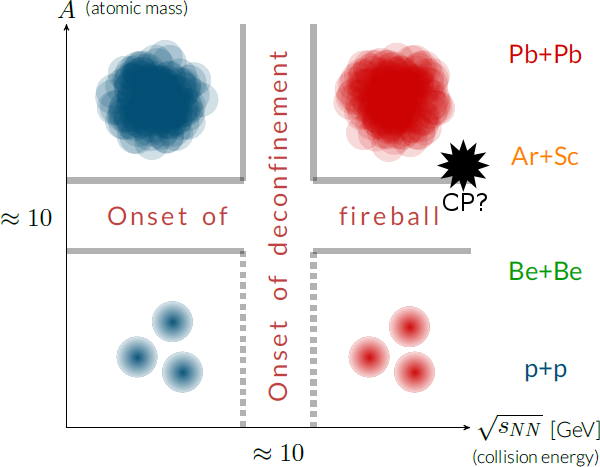}
\end{minipage}
\hspace{0.02\linewidth}
\begin{minipage}{0.64\linewidth}
\caption{The four domains of hadron production defined by the onset of deconfinement and onset of fireball.}
\label{fig7}
\end{minipage}
\end{figure}

``Horn'' and ``step'' in particle ratio and inverse slope are predicted as signature of onset of deconfinement.
They are now also appearing in lighter systems as p+p/Be+Be.
However surprisingly, Be+Be behaves similarly to p+p, while Ar+Sc is between p+p/Be+Be and Pb+Pb.
This may indicate a ``second'' onset: of fireball.
Ar+Sc/Xe+La are still being analysed; it is expected that these will provide further information.
The two onsets may indicates four domains of hadron production separated by two thresholds: onset of deconfinement and onset of fireball, as shown in Figure~\ref{fig7}.
While onset of deconfinement is well established for central Pb+Pb collisions, it is questionable for p+p and light nuclei.

\end{document}